\documentclass{ws-procs9x6adap}
\usepackage{graphicx}

\usepackage{amssymb}

\def\0{\over } \def\2{{\textstyle{1\over2}}} \def\4{{\textstyle{1\over4}}}
\def\5{\hat } \def\6{\partial }

\newcommand{\be}{\begin{equation}}
\newcommand{\ee}{\end{equation}}

\newcommand{\bea}{\begin{eqnarray}}
\newcommand{\eea}{\end{eqnarray}}

\newcommand{\nn}{\nonumber\\ }

\def\Tr{{\,\rm Tr\,}}
\def\Im{{\,\rm Im\,}}
\def\Re{{\,\rm Re\,}}
\def\tr{{\,\rm tr\,}}

\def\CL{\mathcal L }

\begin{document}
\title{
HTL-%
resummed thermodynamics of\\
hot and dense QCD: An update}
\author{Anton Rebhan}
\address{Institut f\"ur Theoretische Physik, Technische Universit\"at Wien,\\
Wiedner Hauptstra\ss e 8--10, 1040 Vienna, Austria}
\maketitle

\begin{picture}(0,0)
\put(263,170){\tt TUW-03-02}
\end{picture}

\abstracts{
We review the proposal to resum the physics of
hard thermal loops in the thermodynamics of the quark-gluon
plasma through
nonperturbative expressions for entropy and density
obtained from a $\Phi$-derivable two-loop approximation.
A comparison with the recently solved large-$N_f$ limit of hot QCD
is performed, and some updates, in particular
on quark number susceptibilities, are made.
}

\section{Introduction}

\begin{figure}
\centerline{\includegraphics[bb=70 240 540 550,width=2.2in]{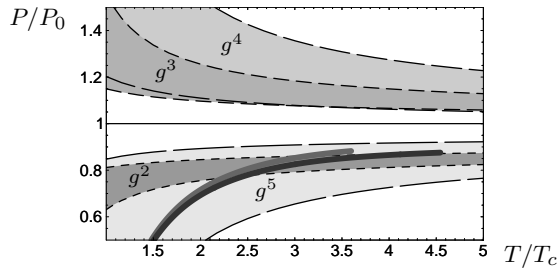}
\begin{picture}(0,0)
\put(-185,95){\small $P/P_0$}
\put(2,5){\small $T/T_c$}
\put(-92,30){\footnotesize $g^5$}
\put(-105,85){\footnotesize $g^4$}
\put(-130,75){\footnotesize $g^3$}
\put(-140,35){\footnotesize $g^2$}
\end{picture}
}
\caption{The poor convergence of thermal perturbation theory in
pure glue QCD. The various grey bands bounded by differently
dashed lines show the perturbative results
to order $g^2$, $g^3$, $g^4$, and $g^5$ with $\overline{\hbox{MS}}$
renormalization
point $\bar\mu$ varied between $\pi T$ and $4\pi T$. The thick
dark-grey line shows the lattice results from the Bielefeld group%
\protect\cite{Boyd:1996bx};
the lighter one behind that of a more recent lattice calculation
using an RG-improved action
from the CP-PACS collaboration\protect\cite{Okamoto:1999hi}.
\label{fig:qcd}}
\end{figure}

Even at temperatures several orders of 
magnitude higher than $\Lambda_{\rm QCD}$
the perturbative series for the thermodynamic potentials of hot QCD\cite{%
Arnold:1995eb,Zhai:1995ac,Braaten:1996jr}
does not appear to converge 
and thus seems to be devoid
of predictive power. While the first correction at order
$g^2$ gives a reasonable estimate for the results obtained in
lattice gauge theory a few times above the transition temperature,
everything breaks down as soon as collective phenomena such as
Debye screening come into the play and produce formally 
higher-order contributions
suppressed by only single powers of $g$ (see Fig.~\ref{fig:qcd}).

Since similar difficulties have been observed in simple scalar
models\cite{Parwani:1995zz,Drummond:1997cw}, the reason for this
failure does not have to do so much
with the fact that QCD has a nonperturbative
sector even in its deconfined phase which limits the number of
computable perturbative coefficients. It rather seems that
screening effects should better not be treated in a strictly
perturbative way. A first encouraging attempt for improving the situation was
put forward by Karsch et al.\cite{Karsch:1997gj}, who proposed
to keep a screening mass unexpanded at any given order of the
loop expansion and to fix this mass by a stationarity principle.
Technically, this corresponds to rewriting the Lagrangian as
\be
\CL=\underbrace{\CL_0-\2 m^2\phi^2}_{\CL_0'}+
\underbrace{\CL_{\rm int}+\2 m^2\phi^2}_{\CL_{\rm int}'}.
\ee

This 
optimization of thermal perturbation theory was
successfully applied to $\phi^4$ theory to three-loop order%
\cite{Andersen:2000yj} and
extended to QCD by Andersen et al.\cite{Andersen:1999,
Andersen:2002ey}. There they proposed replacing the simple
mass term by the gauge-invariant hard-thermal-loop (HTL)
action\cite{Braaten:1992gm,Frenkel:1992ts}. This method, which
they termed HTL perturbation theory (HTLPT), is in principle
a systematic and manifestly gauge invariant scheme. From a
physical point of view, it has, however, the somewhat unsatisfactory
feature that HTL's are used uniformly for soft and hard
momenta although the HTL effective action is accurate only for
soft momenta (and soft virtuality) $\ll T$. 
In the following we shall aim at a resummation that
captures more closely the physics of the collective excitations of
the quark-gluon plasma.

\section{$\Phi$-derivable approximations and HTL resummation} 

Our proposal\cite{%
Blaizot:1999ip,Blaizot:1999ap,Blaizot:2000fc} is
to implement HTL resummations in the spirit of the so-called
$\Phi$-derivable approximations\cite{Baym:1962} of the 2PI skeleton expansion.
In the latter, the thermodynamic potential is expressed in
terms of dressed propagators ($D$ for bosons, $S$ for fermions)
according to
\bea\label{LWQCD}
\Omega[D,S]&=&\2 T \Tr \log D^{-1}-\2 T \Tr \Pi D\nn&&
- T \Tr \log S^{-1} + T\Tr \Sigma S + T\Phi[D,S] 
\eea
where 
$\Phi[D,S]$ is the sum of 2-particle-irreducible ``skeleton''
diagrams. The self-energies $\Pi=D^{-1}-D^{-1}_0$ and 
$\Sigma=S^{-1}-S^{-1}_0$,
where $D_0$ and $S_0$ are bare propagators, are themselves functionals
of the full propagators, determined by the stationarity property
\be\label{staty}
{\delta \Omega[D,S]/\delta D}=0={\delta \Omega[D,S]/\delta S}.
\ee
according to
\be\label{selfenergies}
{\delta\Phi[D,S]/\delta D}=\2\Pi,\quad
{\delta\Phi[D,S]/\delta S}=\Sigma.
\ee

The $\Phi$-derivable two-loop approximation consists of keeping only
the two-loop skeleton diagrams, which leads to
a dressed one-loop approximation for the self-energies
(\ref{selfenergies}). In
a gauge theory this introduces gauge dependences (which are however
parametrically suppressed\cite{Arrizabalaga:2002hn}), but we shall
construct further approximations which are manifestly gauge independent.

A self-consistent
two-loop approximation for $\Omega$ has a remarkable consequence
for the first derivatives of the thermodynamic potential, the entropy
and the number densities:
\be\label{SNdef}
{\mathcal S}=-{\6(\Omega/V)\0\6T}\Big|_{\mu},\quad
{\mathcal N}=-{\6(\Omega/V)\0\6\mu}\Big|_{T}.
\ee
Because of the stationarity property (\ref{staty}), one can ignore the $T$ and
$\mu$ dependences implicit in the spectral densities of the full
propagators, and differentiate exclusively the statistical
distribution functions $n$ and $f$ in (\ref{LWQCD}).
Now the derivative of the {\em two-loop} functional $T\Phi[D,S]$ at fixed
spectral densities of the propagators $D$ and $S$ 
turns out to cancel
part of the terms $\Im(\Pi D)$ and $\Im(\Sigma S)$ in
(\ref{LWQCD}) leading to
the remarkably simple 
formulae\cite{Vanderheyden:1998ph,Blaizot:1999ap,Blaizot:2000fc}
\bea
\label{S2loop}
{\mathcal S}&=&-\tr \int{d^4k\0(2\pi)^4}{\6n(\omega)\0\6T} \left[ \Im 
\log D^{-1}-\Im \Pi \Re D \right] \nn
&&-2\tr \int{d^4k\0(2\pi)^4}{\6f(\omega)\0\6T} \left[ \Im
\log S^{-1}-\Im \Sigma \Re S \right], \\
\label{N2loop}
{\mathcal N}&=&-2\tr \int{d^4k\0(2\pi)^4}{\6f(\omega)\0\6\mu} \left[ \Im
\log S^{-1}-\Im \Sigma \Re S \right].
\eea

Through this formulae, all interactions below order $g^4$ are
summarized by spectral data only, which shows that entropy and density
are the preferred quantities for a quasiparticle description.

The leading-order interaction terms $\propto g^2$ arise from
hard loop momenta involving only the light-cone projection of
the self-energies, e.g.~in pure glue QCD
\be\label{SG2}
{\mathcal S}_2 
 =2N_g\int\!\!{d^4k\0(2\pi)^4}\,{\6n\0\6T}\,\Re{\Pi_T}\!\!\!
\underbrace{\Im\frac{1}{\omega^2-k^2}}_{-
\pi\epsilon(\omega)\delta(\omega^2-k^2)},\ee
where $\Pi_T$ is the transverse component of the gluon self-energy,
which on the light-cone is accurately (to order $g^2$) given
by its HTL value $\5\Pi_T(k,k)=\5 m_\infty^2=\2 \5m_D^2$
even though $k$ is no longer soft\cite{Kraemmer:1990dr,Flechsig:1996ju}.

The more critical $g^3$ (``plasmon'') 
term in the thermodynamic potentials, on the
other hand, arises in an unusual manner: while in the pressure it is
determined by the soft momentum regime of the one-loop contribution,
in the above expressions for the entropy, there are two distinct
origins. One part comes from order-$g$ corrections to
$\Pi_T(k,k)$ at hard momenta $k\sim T$ and are to be interpreted
as a correction to the entropy of hard gluons. Only a fraction ($<1/4$)
arises as the entropy of soft gluons in the HTL 
approximation\cite{Blaizot:1999ip}.
%
The required correction for the hard excitations is
\be
\delta m_\infty^2(k)=\Re \delta\Pi_T(k,k)
=\Re(\begin{picture}(0,0)(0,0)
\put(25,0){\small +}
\put(56,0){\small +}
\put(104,0){\small +}
\end{picture}
\!\!\includegraphics[bb=145 430 500 465,width=5.5cm]{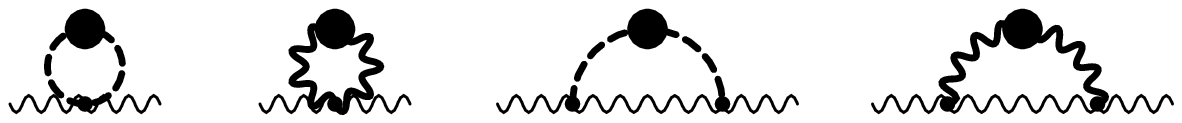})
\ee
and this can be calculated by standard\cite{Braaten:1990mz} 
HTL perturbation theory\footnote{Because
the external momentum is hard, no HTL vertices are needed, and only
one propagator has to be dressed by HTL.}.

These corrections to the asymptotic thermal masses are, in contrast
to the latter, nontrivial functions of the momentum, and can
be evaluated only numerically. However, as far as the generation
of the plasmon term is concerned, these functions contribute
in a certain averaged form which can be calculated analytically,
\be\label{deltamas}
\bar\delta m_\infty^2={\int dk\,k\,n'(k) \Re \delta\Pi_T(\omega=k) 
\0 \int dk\,k\,n'(k)}=-{1\02\pi}g^2NT\hat m_D.
\ee

This result pertains only to the hard excitations;
corrections to the various thermal masses of soft excitations
are known to differ substantially from (\ref{deltamas}).
For instance, the relative correction to the gluonic
plasma frequency\cite{Schulz:1994gf} at $k=0$,
$\delta m^2_{pl.}/\hat m^2_{pl.}$, 
is only about a third of 
$\bar\delta m_\infty^2/m_\infty^2$; the NLO correction to the
nonabelian Debye mass on the other hand is even positive
for small coupling and moreover logarithmically enhanced%
\cite{Rebhan:1993az}.

For an estimate of the effects of a proper incorporation of the
next-to-leading order corrections we have therefore proposed
to include the latter only for hard excitations and to
define our next-to-leading approximation (for gluons) through
$
{\mathcal S}_{NLA}={\mathcal S}_{HTL}\big|_{\rm soft}+
{\mathcal S}_{\bar m_\infty^2}\big|_{\rm hard},
$
where $\bar m_\infty^2$ includes (\ref{deltamas})
and where ${\mathcal S}_{HTL}$ refers to evaluating the 
$\Phi$-derivable 2-loop entropy in the HTL approximation
exactly.
To separate soft ($k\sim \hat m_D$) and hard ($k \sim 2\pi T$) momentum
scales, we introduce the intermediate scale
$\Lambda=\sqrt{2\pi T\hat m_D c_\Lambda}$
and consider a variation of $c_\Lambda=\2\ldots 2$ as part of
our theoretical error estimate. 

Another crucial issue concerns the definition of the corrected
asymptotic mass $\bar m_\infty$. For the range of coupling constants
of interest ($g> 1$), the correction $ |\bar\delta m_\infty^2|$
is greater than the LO value $m_\infty^2$, leading to tachyonic
masses if included in a strictly perturbative manner.

However, this problem is not at all specific to QCD. In the
simple $g^2\varphi^4$ model, one-loop resummed perturbation theory
gives 
\be\label{mstrpert}
m^2=g^2T^2(1-{3\0\pi}g)
\ee 
which also turns tachyonic
for $g>1$. On the other hand, the solution of the 
one-loop gap equation is a monotonic function in $g$, and
it turns out that the first two terms in a $(m/T)$-expansion of
this gap equation,
\be\label{mtruncgap}
m^2=g^2T^2-{3\0\pi}g^2Tm,
\ee
which is perturbatively equivalent to (\ref{mstrpert}), has
a solution that is very 
close to that of the full gap
equation (for $\overline{\mbox{MS}}$ renormalization scale $\bar\mu \approx 2\pi T$)\cite{Blaizot:2000fc}.

In QCD, where the non-local gap equations are too complicated to
be attacked directly we adopted (\ref{mtruncgap}) as a model
to include $\bar\delta m_\infty^2$, which is needed for
the completion of the plasmon effect.\footnote{Numerically,
this differs only slightly from the Pad\'e approximants that
we employed in our first publications%
\cite{Blaizot:1999ip,Blaizot:1999ap}.} This leads to
\be\label{minftygap}
\bar m_\infty^2 = {\textstyle{1\06}}(N+\2 N_f)g^2T^2
-{1\0\sqrt2 \pi}g^2NT\bar m_\infty.
\ee

Up to a single integration constant the resulting entropy expression
determines the thermodynamic pressure. Choosing this (strictly
nonperturbative\cite{Blaizot:1999ip}) input such that e.g.~%
$P(T_c)\approx0$, where $T_c$ is taken from the lattice, this
ambiguity in fact becomes quickly negligible for larger $T/T_c$,
because the contribution of the
bag constant thus introduced drops like $T^{-4}$ in
the normalized pressure $P/P_0$, where $P_0$ is the ideal-gas limit.

The main uncertainty rather comes from the choice of the
renormalization point $\bar\mu$. 
In the following we 
always consider varying $\bar\mu$
by a factor of 2 around a central value of $2\pi T$
and determine the strong coupling constant from the
2-loop renormalization group equation.

\begin{figure}
\centerline{\includegraphics[bb=70 220 540 550,width=3in]{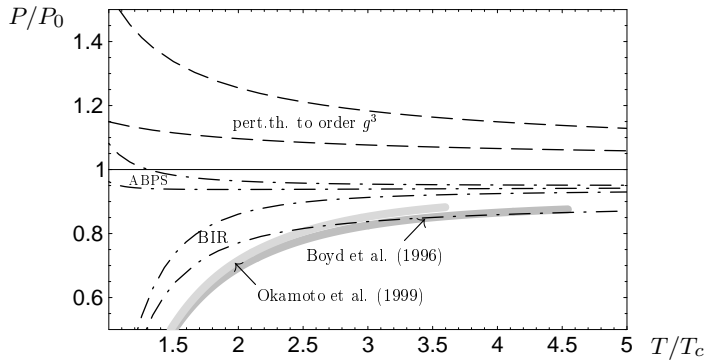}
\begin{picture}(0,0)
\put(3,2){\small $T/T_c$}
\put(-240,127){\small $P/P_0$}
\end{picture}
}
\caption{The pressure in pure-glue QCD in the next-to-leading
approximation developed in Ref.~\protect\cite{Blaizot:2000fc} (BIR)
compared to perturbation theory through order $g^3$ and to
the recent 2-loop HTLPT result\protect\cite{Andersen:2002ey} (ABPS), all
with $\bar\mu$ varied between $\pi T$ and $4\pi T$. The
gray bands represent (continuum extrapolated) lattice 
results\protect\cite{Boyd:1996bx,Okamoto:1999hi}.
\label{fig:p}}
\end{figure}

The result of this procedure is displayed in Fig.~\ref{fig:p}
and compared with lattice data from the Bielefeld\cite{Boyd:1996bx}
and PC-PACS\cite{Okamoto:1999hi} groups, and also with
the recent two-loop calculation in HTLPT\cite{Andersen:2002ey}.
This shows a clear improvement compared to the perturbative result
to order $g^3$, and remarkable agreement with lattice data for
$T\gtrsim 3 T_c$. 

Compared to the HTLPT calculation, an important difference of
our approach is 
the separate treatment of hard and soft contributions, but
the HTLPT result also has a 
large
$g^5$-contribution\cite{Andersen:2002ey} 
with 
opposite sign from that obtained
in 3-loop perturbation theory.

\section{Inclusion of fermions and the large-$N_f$ limit}

When fermions are included ($N_f\not=0$), part of the plasmon
effect in the pressure (and all of the plasmon effect in the
fermion density) is contributed by next-to-leading order
corrections to the asymptotic thermal mass of the fermions,
whose leading-order (HTL) value is $\hat M_\infty^2=g^2 C_f T^2/4$,
with $C_f=2N/N_g$. These can be calculated in standard HTL
perturbation theory according to
\be
{
{1\02k}}\delta M_\infty^2(k)=\delta\Sigma_+(\omega=k) 
=\Re(\begin{picture}(0,0)(0,0)
\put(42,0){\small +}
\end{picture}
\includegraphics[bb=75 435 285 475,width=3.2cm]{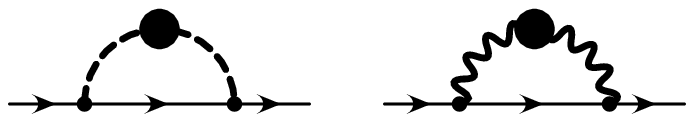})
|_{\omega=k}\;,
\ee
which is again a function of $k$ that can be evaluated only
numerically. However in the plasmon effect it enters only in the
analytically calculable averaged form
\be\label{deltaMas}
\bar\delta M_\infty^2={\int dk\,k\,f'(k) \Re 
2k \delta\Sigma_+(\omega=k) 
\0 \int dk\,k\,f'(k)}=-{1\02\pi}g^2C_fT\hat m_D\;.
\ee

In our previous work\cite{Blaizot:2000fc} we have incorporated
this correction in complete analogy to the gluonic asymptotic mass
(\ref{minftygap}), that is, we have adopted a gap
equation quadratic in $\bar M_\infty$ 
that is perturbatively equivalent to (\ref{deltaMas}).

However, the recent work on the 
large-$N_f$ limit of QCD by Moore\cite{Moore:2002md} has triggered
us to reconsider this procedure, because it turns out that a
quadratic gap equation for the fermions does 
not comply with the large-$N_f$
limit. 

In the large-$N_f$ limit, $N_f\to\infty$, $g^2\to0$ such that 
$g^2_{\rm eff.}=g^2N_f/2\sim 1$,
the quadratic gap equation for the gluons has the correct behaviour
that $\bar m_\infty^2\to g^2_{\rm eff.}T^2/6 + O(1/N_f)$.

Fermion self-energies are suppressed by a factor $1/N_f$, but
we still need to consider them in our expressions for entropy and
density because there are $N_f$ fermions which together produce
a $N_f^0$ contribution to the fermionic entropy. This
precisely equals $-NN_f \bar M_\infty^2 T/6$, where $\bar M_\infty^2$
represents the average appearing in (\ref{deltaMas}). 
The latter involves the correction term calculated in (\ref{deltaMas}),
but without further (rainbow-like) corrections on the internal fermion line.
The fermionic ``gap equation'' thus has to remain linear
in $\bar M_\infty^2$. An equation for $\bar M_\infty^2$ with
the correct behaviour in the large-$N_f$ limit is given by
\be\label{newMgap}
\bar M_\infty^2=g^2 C_f T^2/4 - {1\0\sqrt2 \pi}g^2 C_f T \bar m_\infty.
\ee 
There is then no negative feedback from the fermion mass itself, it
only inherits higher-order terms from the solution to $\bar m_\infty$
(when $N_f$ is finite).

\begin{figure}
\vspace{-5pt}
\begin{minipage}[t]{.47\linewidth}
\centerline{\includegraphics[bb=70 230 540 560,width=\linewidth]{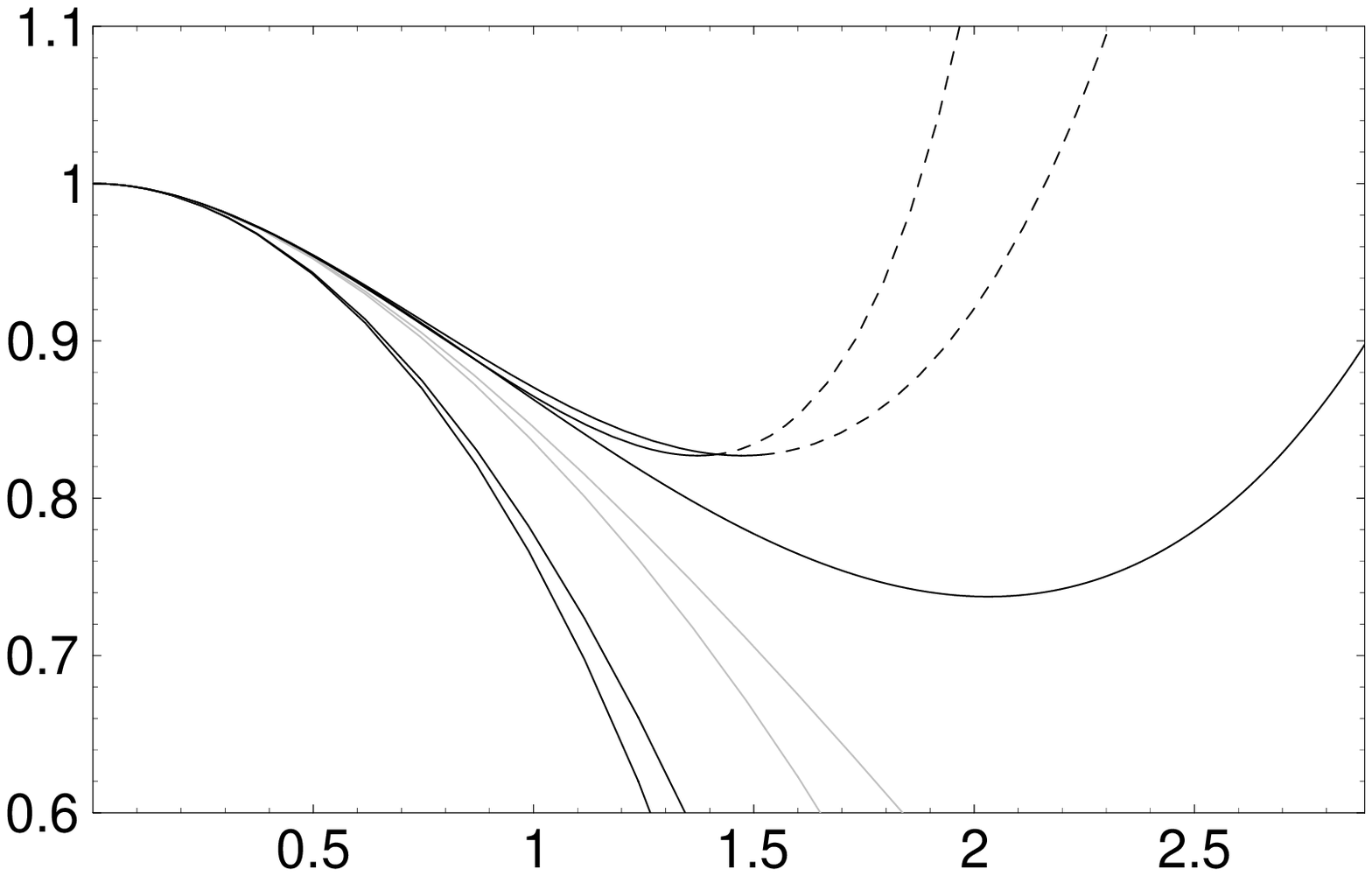}
\begin{picture}(0,0)
\put(-150,107){\footnotesize ${P/P_0}$ (large-$N_f$ limit)}
\put(-25,0){\footnotesize $\hat m_D/T$}
\put(-36,50){\tiny exact --}
\end{picture}
}
\caption{The exact result\protect\cite{Ipp:2003zr} 
for the pressure in the limit of
large $N_f$ compared with the $\Phi$-derivable 2-loop result
in the HTL approximation (full lines) and in the next-to-leading
approximation (full lines ending in dashed lines), for $\bar\mu=T$
and $4\pi T$. The gray lines denote the next-to-leading
approximation with quadratic fermionic gap equation considered
in Ref.~\protect\cite{Blaizot:2000fc}, but which we argue needs to be
replaced by Eq.~(\protect\ref{newMgap}).
\label{aicomp}}
\end{minipage}\hfill
\begin{minipage}[t]{.47\linewidth}
\centerline{\includegraphics[bb=93 243 543 573,width=\linewidth]{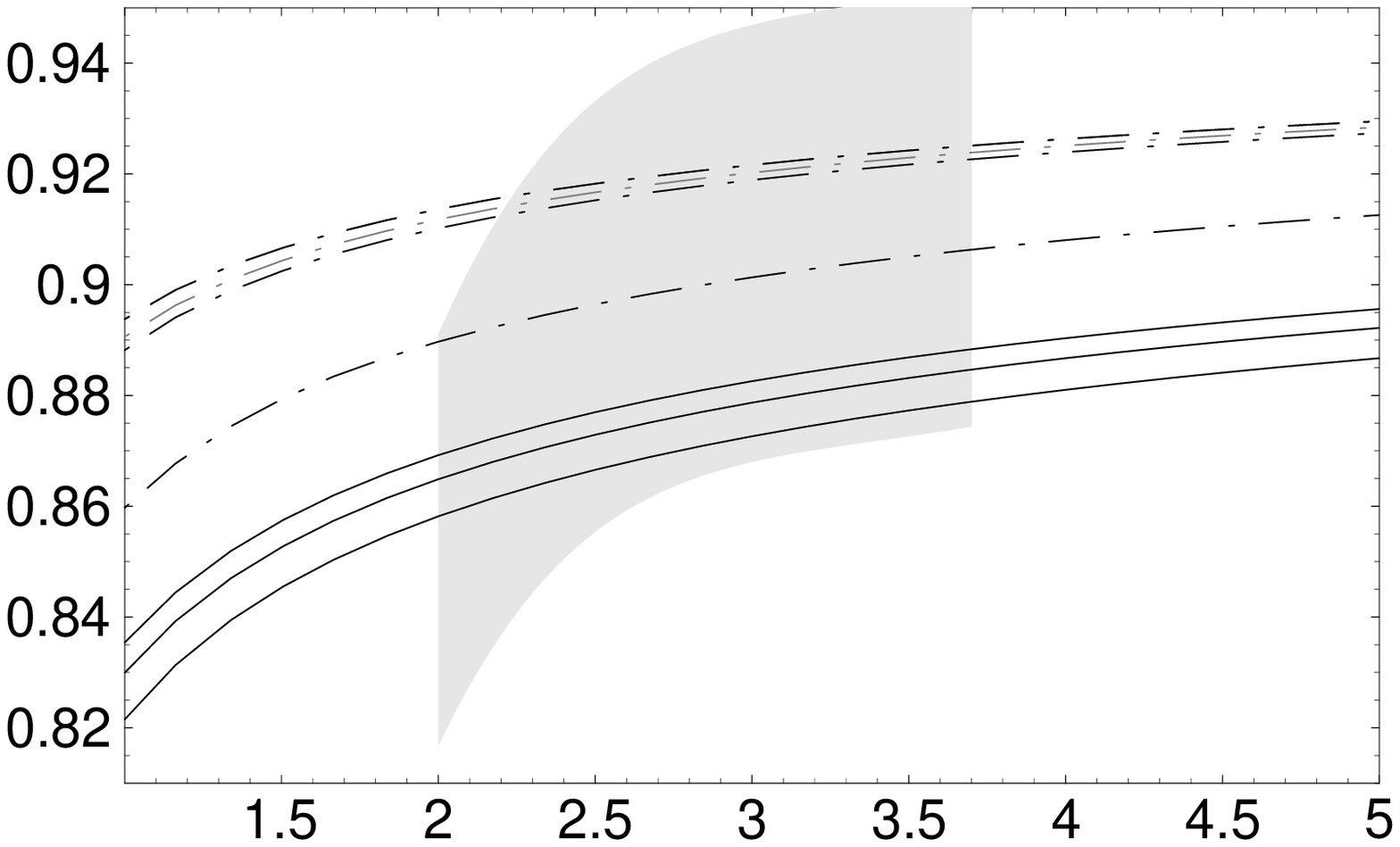}
\begin{picture}(0,0)
\put(-150,107){\footnotesize ${\mathcal S/\mathcal S_0}$ ($N_f=0,2,3$)}
\put(-23,0){\footnotesize $T/T_c$}
\end{picture}
}
\caption{The entropy in the HTL approximation (full lines, $N_f=0,3,2$
from bottom to top) and in the next-to-leading approximation
(dash-dotted lines, $N_f=0,2,3$ from bottom to top) with $\bar\mu$=$2\pi T$
and compared with an estimated continuum 
extrapolation\protect\cite{Karsch:1999vy} 
of the
lattice result for $N_f$=2.
Only the NLA $N_f$=2 result changes by switching to the
fermionic gap equation (\protect\ref{newMgap}) from the gray dash-dotted
line to the black one just below.
\label{fig:SNf}}
\end{minipage}\hfill
\end{figure}

In Fig.~\ref{aicomp} we compare the 2-loop $\Phi$-derivable result
in our approximations in the limit of large $N_f$ with the exact result%
\footnote{The result published in Ref.~\citelow{Moore:2002md}
has recently been found to be in error; the corrected exact result
can be found in Ref.~\citelow{Ipp:2003zr}.}. The latter has the
curious behaviour of being nonmonotonic as $g^2_{\rm eff}$ is
increased, and this behaviour is in fact qualitatively reproduced
in our approach with (\ref{newMgap}).
However,
when 
$g_{\rm eff}^2 \gtrapprox 7.4$,
$M_\infty^2$ becomes negative (dashed lines in Fig.~\ref{aicomp}).
This is not a problem
for the $\Phi$-derivable expressions at order $N_f^0$, but it
means that for large but still finite $N_f$ the fermionic quasiparticles
(at least in our approximation)
cease to exist. 

However, in our applications to real QCD the revised fermionic
gap equation (\ref{newMgap}) does not have the problem of giving
rise to tachyonic masses even close to the transition temperature.
If we therefore compare the results of our approximations
in the large $N_f$ limit with the exact one only in the region
where $\bar M_\infty^2$ remains positive, the outcome is
in fact encouraging: the agreement below the point where
$\bar M_\infty^2$ 
vanishes 
is remarkably good
even though the coupling is no longer small
and $\hat m_D/T \equiv g_{\rm eff.}/\sqrt3 \sim 1$.

By a curious coincidence, for $N=3$ and $N_f=3$
the revised gap equation (\ref{newMgap})
together with (\ref{minftygap}) has exactly the same solutions
as the uncoupled quadratic gap equations we used previously.
Only for $N_f>3$ there is at all a coupling where the fermionic
mass ceases to grow monotonicly with $g$; for $N_f\le 3$ this
never happens. Because of this coincidence, the numerical
changes in our previous results are almost completely
negligible. As an example, Fig.~\ref{fig:SNf} updates
the results published previously\cite{Blaizot:2000fc} for the entropy with
flavour numbers $N_f=0,2,3$ in comparison with the estimated
continuum extrapolation of Ref.~\cite{Karsch:1999vy}. Only the NLA
result for $N_f=2$
changes at all from the gray dash-dotted line to the one slightly
below it. 

\begin{figure}
\begin{minipage}[t]{.47\linewidth}
\centerline{\includegraphics[bb=70 240 540 550,width=\linewidth]{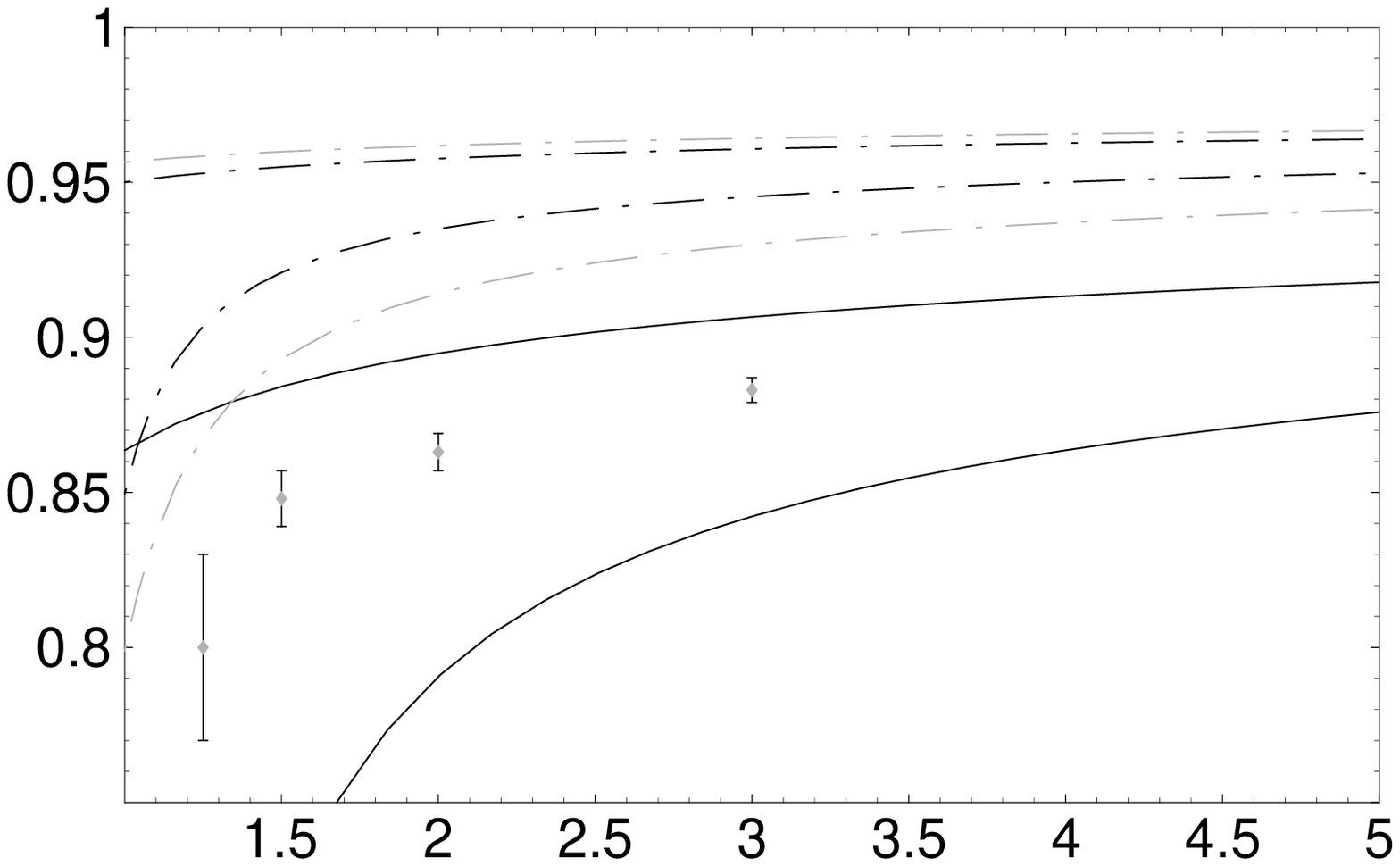}
\begin{picture}(0,0)
\put(-150,102){\footnotesize ${\chi/\chi_0}$}
\put(-20,-2){\footnotesize $T/T_c$}
\end{picture}
}
\caption{
Comparison of our results for $\chi/\chi_0$ in massless $N_f$=2 QCD
with the lattice results of Ref.~\protect\cite{Gavai:2001ie} 
(no continuum extrapolation).
Full lines refer to the HTL approximation, dash-dotted lines to
NLA. (Gray dash-dotted denotes our previous NLA
estimate\protect\cite{Blaizot:2001vr}.)
\label{fig:chinf2}}
\end{minipage}\hfill
\begin{minipage}[t]{.47\linewidth}
\centerline{\includegraphics[bb=70 240 540 550,width=\linewidth]{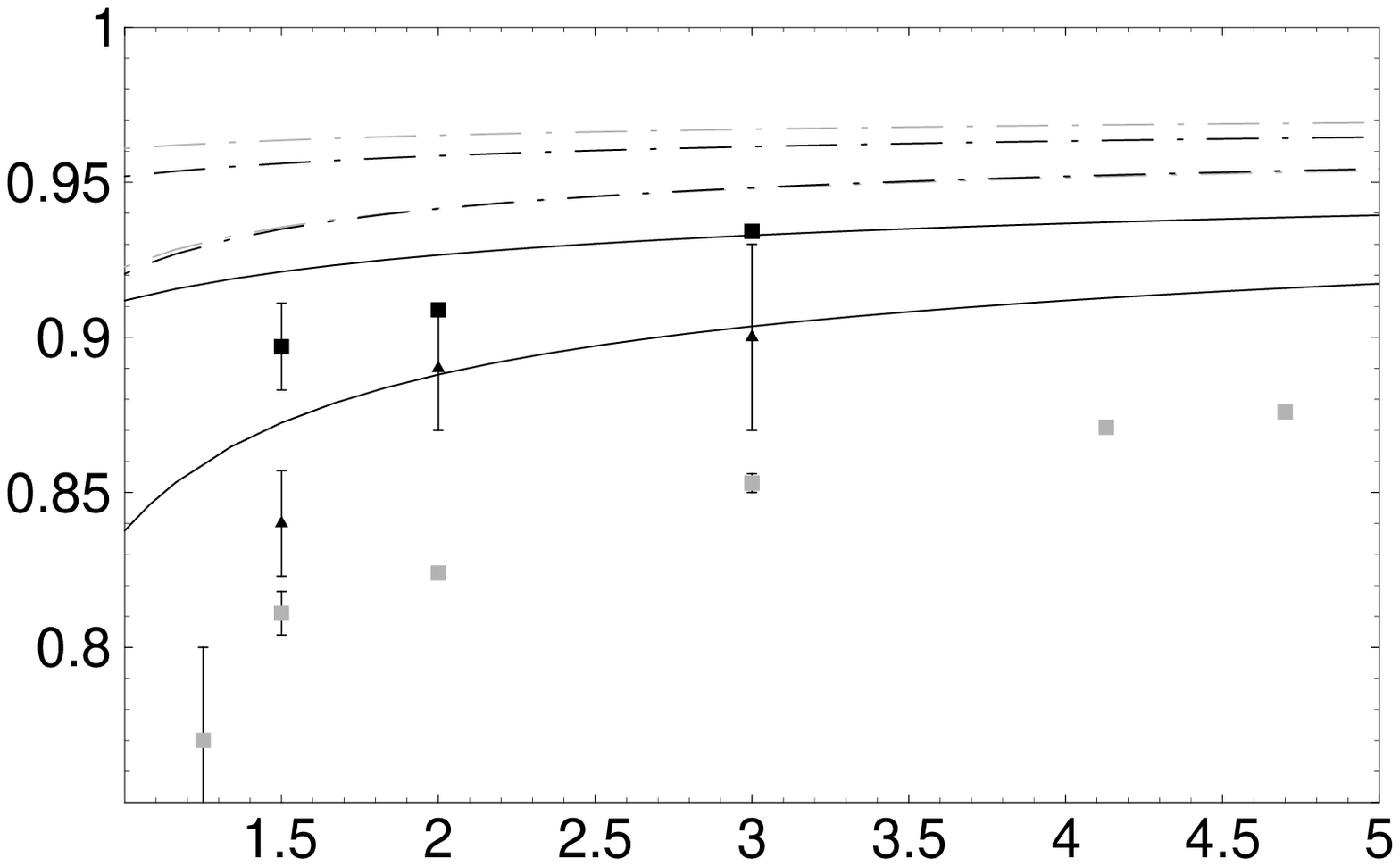}
\begin{picture}(0,0)
\put(-150,102){\footnotesize ${\chi/\chi_0}$}
\put(-20,-2){\footnotesize $T/T_c$}
\end{picture}
}
\caption{Comparison of our updated results for $\chi/\chi_0$ 
in the formal limit $N_f$=$0$ 
with the previous
lattice results for quenched QCD of Ref.~\protect\cite{Gavai:2001fr}
(gray data points) and two recent continuum 
extrapolations\protect\cite{Gavai:2002jt}
(black data points).
\label{fig:chinf0}}
\end{minipage}\hfill
\end{figure}

More interestingly, with the new fermionic gap equation (\ref{newMgap})
we can narrow down somewhat 
our predictions\cite{Blaizot:2001vr} 
for the quark number susceptibilities.
Previously we have
determined our estimated theoretical errors 
for the latter by combining the results obtained
from a quadratic gap equation with the results from a Pad\'e
approximant. To be compatible with the large-$N_f$ limit, we
can now restrict to (\ref{newMgap}) and produce somewhat narrower
error bands (which still are dominated by the 
$\bar\mu$ dependence). Fig.~\ref{fig:chinf2} shows our prediction
for $N_f=2$ in comparison with available lattice data\cite{Gavai:2001ie}.
Fig.~\ref{fig:chinf0} shows the results for quenched QCD
together with recent results for two different continuum
extrapolations\cite{Gavai:2002jt} (both are higher than the previous lattice 
results\cite{Gavai:2001fr}).

It would be interesting to compare our predictions with HTLPT to two-loop
order; the one-loop HTL-resummed results from
charge correlators\cite{Chakraborty:2001kx} 
suffer from severe
overcounting while missing out the plasmon term\cite{Blaizot:2002xz} 
so that one should not perform a comparison yet.%
\footnote{For results and conjectures based on the
recent determination of the $g^6\ln(1/g)$ term in the
thermodynamic potential\cite{Kajantie:2002wa} see 
Ref.~\cite{Vuorinen:2002ue}.}


\vspace{-10pt}
\section{Outlook}

The HTL-based quasiparticle description of the thermodynamics of
hot QCD that we have developed can be straightforwardly extended to finite
chemical potential. First steps in this direction 
that go beyond 
the simple quasiparticle models
of Ref.~\cite{Peshier:2002ww} are indeed
encouraging\cite{Romatschke:2002pb}. Further refinements,
in particular a full inclusion of the momentum dependence
of the next-to-leading order asymptotic masses, is work in progress.

\section*{Acknowledgments}

The work presented here has been carried out in collaboration
with Jean-Paul Blaizot and Edmond Iancu, whom I wish to thank
cordially. I would also like to thank Guy Moore for discussions and
Michael Schmidt and the organizers of SEWM02 for all their efforts and
their hospitality.


\begin{thebibliography}{39}
\newcommand{\enquote}[1]{``#1''}

\bibitem{Arnold:1995eb}
P.~Arnold and C.-X. Zhai, {\it Phys. Rev.\/} {\bf D51}, 1906 (1995).

\bibitem{Zhai:1995ac}
C.-X. Zhai and B.~Kastening, {\it Phys. Rev.\/} {\bf D52}, 7232 (1995).

\bibitem{Braaten:1996jr}
E.~Braaten and A.~Nieto, {\it Phys. Rev.\/} {\bf D53}, 3421 (1996).

\bibitem{Boyd:1996bx}
G.~Boyd {\it et~al.\/}, {\it Nucl. Phys.\/} {\bf B469}, 419 (1996).

\bibitem{Okamoto:1999hi}
M.~Okamoto {\it et~al.\/}, {\it Phys. Rev.\/} {\bf D60}, 094510 (1999).

\bibitem{Parwani:1995zz}
R.~Parwani and H.~Singh, {\it Phys. Rev.\/} {\bf D51}, 4518 (1995).

\bibitem{Drummond:1997cw}
I.~T. Drummond, R.~R. Horgan, P.~V. Landshoff and A.~Rebhan, {\it Nucl.
  Phys.\/} {\bf B524}, 579 (1998).

\bibitem{Karsch:1997gj}
F.~Karsch, A.~Patk{\'o}s and P.~Petreczky, {\it Phys. Lett.\/} {\bf B401}, 69
  (1997).

\bibitem{Andersen:2000yj}
J.~O. Andersen, E.~Braaten and M.~Strickland, {\it Phys. Rev.\/} {\bf D63},
  105008 (2001).

\bibitem{Andersen:1999} 
J.~O. Andersen, E.~Braaten and M.~Strickland, {\it Phys. Rev. Lett.\/} {\bf
  83}, 2139 (1999);
 {\it Phys. Rev.\/} {\bf D61},
  014017, 074016 (2000).

\bibitem{Andersen:2002ey}
J.~O. Andersen, E.~Braaten, E.~Petitgirard and M.~Strickland, {\it Phys.
  Rev.\/} {\bf D66}, 085016 (2002); J.~O. Andersen, these proceedings
[hep-ph/0210195].

\bibitem{Braaten:1992gm}
E.~Braaten and R.~D. Pisarski, {\it Phys. Rev.\/} {\bf D45}, 1827 (1992).

\bibitem{Frenkel:1992ts}
J.~Frenkel and J.~C. Taylor, {\it Nucl. Phys.\/} {\bf B374}, 156 (1992).

\bibitem{Blaizot:1999ip}
J.~P. Blaizot, E.~Iancu and A.~Rebhan, {\it Phys. Rev. Lett.\/} {\bf 83}, 2906
  (1999).

\bibitem{Blaizot:1999ap}
J.~P. Blaizot, E.~Iancu and A.~Rebhan, {\it Phys. Lett.\/} {\bf B470}, 181
  (1999).

\bibitem{Blaizot:2000fc}
J.~P. Blaizot, E.~Iancu and A.~Rebhan, {\it Phys. Rev.\/} {\bf D63}, 065003
  (2001).

\bibitem{Baym:1962}
G.~Baym, {\it Phys. Rev.\/} {\bf 127}, 1391 (1962).

\bibitem{Arrizabalaga:2002hn}
A.~Arrizabalaga and J.~Smit, {\it Phys. Rev.\/} {\bf D66}, 065014 (2002);
these proceedings [hep-ph/0301093].

\bibitem{Vanderheyden:1998ph}
B.~Vanderheyden and G.~Baym, {\it J. Stat. Phys.\/} {\bf 93}, 843 (1998).

\bibitem{Kraemmer:1990dr}
U.~Kraemmer, M.~Kreuzer and A.~Rebhan, {\it Ann. Phys.\/} {\bf 201}, 223
  (1990).

\bibitem{Flechsig:1996ju}
F.~Flechsig and A.~K. Rebhan, {\it Nucl. Phys.\/} {\bf B464}, 279 (1996).

\bibitem{Braaten:1990mz}
E.~Braaten and R.~D. Pisarski, {\it Nucl. Phys.\/} {\bf B337}, 569 (1990).

\bibitem{Schulz:1994gf}
H.~Schulz, {\it Nucl. Phys.\/} {\bf B413}, 353 (1994).

\bibitem{Rebhan:1993az}
A.~K. Rebhan, {\it Phys. Rev.\/} {\bf D48}, 3967 (1993).

\bibitem{Moore:2002md}
G.~D. Moore, {\it JHEP\/} {\bf 0210}, 055 (2002).

\bibitem{Ipp:2003zr}
A.~Ipp, G.~D. Moore and A.~Rebhan, 
{\it JHEP\/} {\bf 0301}, 037 (2003).

\bibitem{Karsch:1999vy}
F.~Karsch, {\it Nucl. Phys. Proc. Suppl.\/} {\bf 83}, 14 (2000).

\bibitem{Blaizot:2001vr}
J.~P. Blaizot, E.~Iancu and A.~Rebhan, {\it Phys. Lett.\/} {\bf B523}, 143
  (2001).

\bibitem{Gavai:2001ie}
R.~V. Gavai, S.~Gupta and P.~Majumdar, {\it Phys. Rev.\/} {\bf D65}, 054506
  (2002).

\bibitem{Gavai:2002jt}
R.~V. Gavai and S.~Gupta, 
hep-lat/0211015.

\bibitem{Gavai:2001fr}
R.~V. Gavai and S.~Gupta, {\it Phys. Rev.\/} {\bf D64}, 074506 (2001).

\bibitem{Chakraborty:2001kx}
P.~Chakraborty, M.~G. Mustafa and M.~H. Thoma, {\it Eur. Phys. J.\/} {\bf C23},
  591 (2002).

\bibitem{Blaizot:2002xz}
J.~P. Blaizot, E.~Iancu and A.~Rebhan, 
{\it Eur. Phys. J.\/} {\bf C27}, 433
  (2003).

\bibitem{Kajantie:2002wa}
K.~Kajantie, M.~Laine, K.~Rummukainen and Y.~Schr{\"o}der, 
hep-ph/0211321.

\bibitem{Vuorinen:2002ue}
A.~Vuorinen, 
hep-ph/0212283.

\bibitem{Peshier:2002ww}
A.~Peshier, B.~K{\"a}mpfer and G.~Soff, {\it Phys. Rev.\/} {\bf D66}, 094003
  (2002).

\bibitem{Romatschke:2002pb}
P.~Romatschke, these proceedings
[hep-ph/0210331].

\end{thebibliography}

\end{document}